\documentclass[11pt]{article}

\usepackage[margin=1in]{geometry}
\usepackage{amsmath}
\usepackage{booktabs}
\usepackage{graphicx}
\usepackage[hidelinks]{hyperref}
\usepackage{placeins}

\newcommand{\Description}[1]{}

\title{Towards Generalizable and Efficient Large-Scale Generative Recommenders}

\author{
Qiuling Xu \quad Ko-Jen Hsiao \quad Moumita Bhattacharya\\
Netflix Research\\
\texttt{\{qiulingx,khsiao,mbhattacharya\}@netflix.com}
}

\date{}

\begin{document}

\maketitle

\begin{abstract}
Generative recommendation models can model user behavior as sequences of events and provide a shared backbone for multiple recommendation tasks. In production, however, pre-training gains do not automatically translate into downstream application improvements: task headroom, repeated-training cost, serving latency, and item freshness all affect transfer. We describe our experience scaling a generative recommender from 2M to 1B backbone parameters, excluding embedding and decoding layers, in a production-scale title recommendation setting. Across multiple downstream tasks, we observe task-dependent scaling behavior: some tasks approach an empirical ceiling within the observed scale range, while others continue to benefit from additional capacity. This motivates using offset scaling-law fits as a diagnostic for where additional model scale may be more or less useful.

We then study production constraints that arise when applying the model in practice. Frequent retraining over trillions of behavior tokens makes training and decoding efficiency important; cached serving can make the immediate next-token target stale; and newly launched titles may need to be scored from semantic metadata before collaborative ID embeddings are reliable. We address these issues with multi-token prediction for serving-latency alignment, sampled softmax and a projected decoding head for efficient repeated training, and semantic item towers with collaborative-embedding masking for cold-start adaptation. In a one-week production-shadow evaluation over 1M users, the 1B-backbone model achieves higher MRR than the 2M-backbone baseline across all reported tasks, including a 22.5\% relative gain on the lower-predictability Task A. Overall, the results support treating model scale as one component of a production transfer problem, alongside task headroom, decoding cost, serving-latency alignment, and item generalization.
\end{abstract}

\vspace{0.5em}

\begin{figure}[!t]
  \centering
  \includegraphics[width=0.95\textwidth]{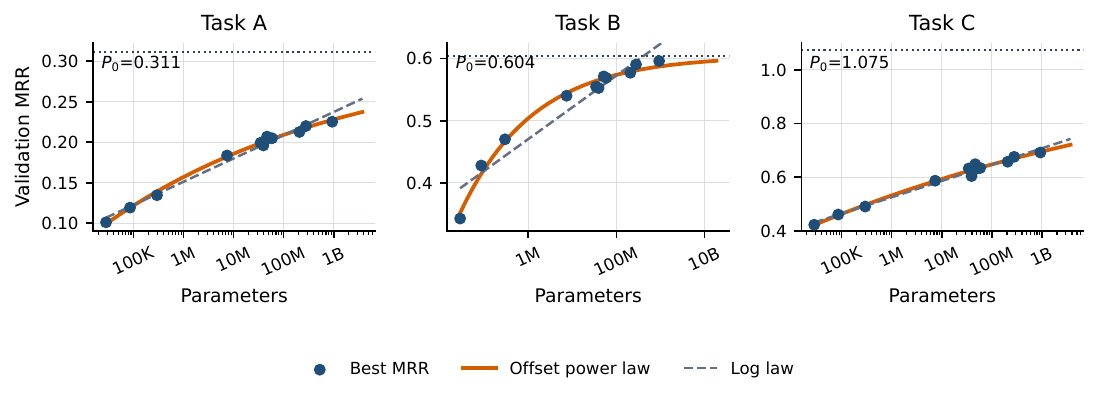}
  \caption{Scaling-law fits for three anonymized recommendation task categories: Task A captures lower-predictability long-horizon taste, Task B captures moderate-predictability short-horizon engagement, and Task C captures higher-predictability time- or availability-driven behavior. The x-axis counts backbone parameters, excluding embedding and decoding layers; Section~\ref{sec:scaling-law} details the fitting setup.}
  \Description{Three plots showing best validation MRR against backbone parameter count for Task A, Task B, and Task C. Task A is lower-predictability long-horizon taste, Task B is moderate-predictability short-horizon engagement, and Task C is higher-predictability time- or availability-driven behavior. Each plot compares observed best MRR points, an offset power-law fit, a log-law fit, and an estimated ceiling P0.}
  \label{fig:scaling-law-examples}
\end{figure}

\section{Introduction}

Generative recommendation models extend the sequence-modeling paradigm of language models to user behavior: a user's history is represented as a sequence of events, and the model is trained to predict future events. This formulation is appealing for industrial recommendation because a single backbone can support many downstream tasks, including retrieval, ranking features, user embeddings, and cold-start scoring. However, simply increasing model size does not by itself guarantee better production recommendation quality. The benefit of scale depends on which tasks still have headroom, whether the model can be retrained frequently enough, and whether the training objective matches how recommendations are served.

In this paper, we study these questions through the development of a large-scale generative recommender trained on trillions of behavior tokens and scaled to 1B backbone parameters, excluding embedding and decoding layers. Our main finding is that scaling is useful but uneven: different recommendation tasks follow different scaling trajectories and have different apparent ceilings. This makes scaling-law analysis a practical tool, not only a diagnostic plot. It helps identify where additional capacity is likely to improve downstream quality and where the bottleneck is more likely to be the objective, serving path, or item representation. The practical value of this analysis is not merely prediction; it helps decide whether the next intervention should be more capacity, a better objective, or a better item representation.

We focus on three production challenges that arise after scaling. First, large models must be efficient enough to retrain repeatedly as user behavior and item catalogs evolve. We use sampled softmax and a shared projected decoding head to reduce direct output-scoring cost while keeping the backbone capacity fixed. Second, many production recommenders rely on cached embeddings, candidates, or scores, which means the immediate next event used in next-token prediction may no longer be the right serving-time target. We address this mismatch with multi-token prediction, which supervises the model on a weighted set of future high-value targets. Third, newly launched titles may not have reliable collaborative embeddings after an extensive training period starts. We combine semantic item towers with collaborative-embedding masking so the model can score titles from content and metadata when ID-based signals are sparse.

The resulting system shows consistent gains in production-shadow evaluation. Compared with the 2M-backbone baseline, the 1B-backbone model improves MRR across all reported Netflix recommendation tasks in a one-week evaluation over 1M users, with a substantial gain on the most challenging task in our evaluation suite. Overall, this paper contributes an industry case study of how to make scale transfer in a production generative recommender: task-dependent offset scaling laws identify where capacity still has headroom; sampled softmax and a projected decoding head reduce repeated-training cost; multi-token prediction aligns supervision with cached serving; and semantic item towers with collaborative-embedding masking improve scoring for unseen or weak-ID titles.

\section{Related Work}

\paragraph{Generative recommenders.}
Generative retrieval casts recommendation as sequence generation over item or semantic identifiers, providing an alternative to nearest-neighbor retrieval over fixed item embeddings. TIGER introduced this formulation with semantic IDs (SIDs) for retrieval and unseen-item generalization~\cite{rajput2023recommender}. HSTU extended the idea to high-cardinality streaming recommendation data and showed that generative recommender quality can scale predictably with training compute~\cite{zhai2024actions}. Recent industrial systems further improve this direction through jagged-tensor context parallelism for longer HSTU histories~\cite{dong2025context}, long-sequence transformer engineering~\cite{chai2025longer}, and foundation models for user activity sequences~\cite{chen2025pinfm}. Other generative recommenders study task-specific tokenization and unified search-recommendation settings~\cite{ma2025grace,shi2025gensar}. Our work is complementary to these architecture and tokenization advances: we study how scaling transfers across downstream tasks under repeated retraining, cached serving, and new or weakly represented items.

\paragraph{Scaling laws in recommendation.}
Scaling laws have become a practical tool for predicting the returns of larger models and datasets in language modeling~\cite{hoffmann2022training}. In recommender systems, prior work has examined power-law behavior for ID-based sequential recommenders and CTR models~\cite{zhang2024scalingseq,lai2025scaling}, while HSTU reports scaling behavior for generative recommenders over large compute ranges~\cite{zhai2024actions}. These works establish that recommendation quality can improve with scale, but they do not fully answer which production tasks still have useful headroom. We focus on this question by fitting task-dependent offset scaling laws and using the fitted ceilings to distinguish tasks where additional capacity is likely to help from tasks where objective or serving changes may matter more.

\paragraph{Efficient training and serving at industrial scale.}
Industrial recommender models must be efficient enough to retrain and serve under changing user behavior, item availability, and latency constraints. LONGER and HSTU context parallelism address the cost of modeling long user histories through architecture and distributed training techniques~\cite{chai2025longer,dong2025context}. Embedding offloading scales sparse recommendation models by moving very large embedding tables through host memory and cache-aware sharding~\cite{park2024toward}. For output scoring, sampled softmax, scalable cross-entropy losses, Cut Cross-Entropy, and sub-item inference reduce different costs in large output layers~\cite{khrylchenko2025logq,mezentsev2024scalable,wijmans2025cut,petrov2024efficient}. Cut Cross-Entropy computes exact cross-entropy without materializing the full token-by-vocabulary logit matrix, primarily reducing memory, while sampled softmax reduces the number of logits evaluated. Our efficiency recipe is intentionally simple: 1\% uniformly sampled negatives and a shared projected $d/8$ decoding head reduce direct item-scoring cost while preserving the backbone capacity studied in the scaling-law analysis. This recipe is complementary to exact memory-efficient kernels such as Cut Cross-Entropy and to SID-style output-space reductions.

\paragraph{Semantic representations and cold start.}
SIDs replace or augment random item identifiers with content-derived discrete structure, improving generalization for sparse, long-tail, or unseen items~\cite{rajput2023recommender,singh2024semanticids,zhu2024cost,mei2025semantic}. PinFM also highlights new-item handling as a central challenge for user-activity foundation models~\cite{chen2025pinfm}. Industrial cold-start work has addressed complementary exposure and exploration problems, such as finding appropriate audiences for new items~\cite{wang2025item}. Our contribution targets the complementary model-representation problem: collaborative item embeddings remain available when interaction evidence is reliable, while multimodal semantic towers and collaborative-embedding masking train the same decoder to score titles from content and metadata when ID-based evidence is weak or unavailable.

\paragraph{Serving-aware objectives and reusable representations.}
Production recommenders often reuse expensive model outputs through cached embeddings, candidate sets, or shared user representations. Generalized user-representation systems show that reusable embeddings can support many downstream tasks with independent serving stacks and lower infrastructure cost~\cite{fazelnia2025generalized}. RADAR uses deferred asynchronous retrieval to cache high-quality candidates generated with a more expensive ranking path~\cite{jaspal2025radar}. These systems motivate a serving-aware view of recommendation modeling. Our MTP objective addresses a complementary issue: when cached outputs are consumed after the user's state has moved forward, the immediate next-token label can become stale. We therefore supervise the model with a weighted set of future high-value targets matched to the serving horizon.

\section{Scaling Law in Recommendation}
\label{sec:scaling-law}

We first ask whether larger generative recommenders improve uniformly across downstream tasks. To reduce temporal distribution shift, we group user activities into three broad anonymized task categories that differ in predictability and serving sensitivity. Table~\ref{tab:task-taxonomy} summarizes this taxonomy with non-confidential descriptions rather than exact Netflix surface definitions. This grouping is intentionally coarse: multiple production canvases can map into the same category, while the taxonomy preserves the main industrial distinction between long-term taste, short-term engagement, and time-dependent behavior.

\begin{table}
  \centering
  \caption{Task categories used for scaling-law analysis.}
  \label{tab:task-taxonomy}
  \begin{tabular}{@{}llp{0.47\linewidth}@{}}
    \toprule
    Task & Predictability & Description \\
    \midrule
    Task A & Lower & Long-horizon taste; sparse positives. \\
    Task B & Moderate & Short-horizon engagement; recent context. \\
    Task C & Higher & Time- or availability-driven population signals. \\
    \bottomrule
  \end{tabular}
  \Description{The table groups anonymized recommendation tasks by predictability and non-confidential behavioral description for the scaling-law analysis.}
\end{table}

For ease of explanation, we group broad recommendation tasks into the three general categories shown in Table~\ref{tab:task-taxonomy}. Task A covers long-horizon personalized taste, where positives can be sparse and the target may extend beyond the current session. Task B covers shorter-horizon engagement, where recent actions and session-level intent are more informative. Task C covers behavior where timing, catalog availability, launch dynamics, or broad population demand provide stronger predictive signals. This abstraction preserves the operational distinctions needed for scaling analysis without exposing internal canvas names or traffic allocations.

For the scaling-law study, we use a controlled time-split evaluation setup. Each example contains a user history ending before a cutoff time, and model inference is evaluated against labels from future behavior after that cutoff. We sample task instances from multiple Netflix recommendation canvases and map them into the three task categories in Table~\ref{tab:task-taxonomy}. Across model scales, we keep the training data, evaluation examples, test window, item vocabulary, embedding dimension, and decoding setup fixed. The ablation varies only the backbone size, so differences in Figure~\ref{fig:scaling-law-examples} primarily reflect sequence-model capacity rather than changes in data, embedding size, or evaluation distribution.

Because thresholded metrics can make smooth improvements appear discontinuous~\cite{schaeffer2023emergent}, we use Mean Reciprocal Rank (MRR) as the primary scaling metric. MRR is continuous enough to capture incremental gains but remains bounded above by 1, making it natural to model task-specific ceilings. We observed similar qualitative trends for test loss, Hit Rate, and NDCG.

We fit each task with an offset power law over non-embedding, non-decoding parameters:

\begin{equation}
  P(N) = P_0 - \left(\frac{N}{N_0}\right)^{-a},
  \quad a > 0.
  \label{eq:scaling-law}
\end{equation}

Here, $P(N)$ is validation performance at scale $N$, $P_0$ is the scale-implied saturation level, $N_0$ is a scale parameter, and $a$ controls the rate of improvement. We use $P_0$ as a task-specific scaling diagnostic under the current data, objective, and evaluation distribution, rather than as an absolute measure of task importance. In practice, the value of additional capacity also depends on task importance, downstream coverage, current baseline quality, and marginal scaling cost. For example, Task A has lower absolute MRR than the other tasks, but its fitted $P_0=0.311$ still leaves meaningful relative headroom; if such a task is product-important, the curve provides evidence that further scaling may remain worthwhile. Task B has a fitted $P_0=0.604$, close to the best observed performance in this scale range, suggesting that further scaling alone is likely to provide smaller marginal gains unless the objective or serving setup creates additional headroom. For Task C, the fitted $P_0=1.075$ is effectively close to the MRR upper bound of 1; the small excess above 1 is estimation error, and the fit suggests that this task remains highly predictable under sufficient data and scale. Figure~\ref{fig:scaling-law-examples} plots raw MRR against log-scaled parameters; the offset power law bends toward $P_0$ in this view and becomes linear only when plotting $\log(P_0-P)$ against $\log N$.

\begin{table}
  \centering
  \caption{Fit-error comparison for scaling-law models.}
  \label{tab:scaling-fit-errors}
  \begin{tabular}{@{}lccc@{}}
    \toprule
    Task & Offset RMSE & Log RMSE & Reduction \\
         & ($10^{-3}$) & ($10^{-3}$) & \\
    \midrule
    Task A & 2.80 & 5.43 & 48.4\% \\
    Task B & 8.16 & 21.34 & 61.8\% \\
    Task C & 9.33 & 10.97 & 14.9\% \\
    \bottomrule
  \end{tabular}
  \Description{A table comparing fit RMSE for offset power-law and log-law scaling models. RMSE values are reported in units of 10 to the negative third, and offset power-law RMSE is lower for Task A, Task B, and Task C.}
\end{table}

We compare this fit with the log-linear form $P=a\log(N)+b$ used in prior generative recommendation work~\cite{zhai2024actions}. The log-linear model can match monotonic improvements over a limited scale range, but it cannot represent bounded metrics that approach task-specific ceilings. Table~\ref{tab:scaling-fit-errors} shows that the offset form reduces RMSE for all three tasks in our scaling-law fits.

For this analysis, we count only backbone parameters and exclude embedding and final decoding layers. This follows the scaling-law convention that embeddings often have secondary effects~\cite{hoffmann2022training} and keeps the scaling analysis focused on sequence-model capacity rather than representation or scoring choices. Empirically, including these layers did not materially change the task ordering or the preference for the offset fit.

\section{Efficient Training and Inference}
\label{sec:efficient-training-and-inference}

Efficiency is one of the main differences between a promising large recommender and an operational one. Unlike many language models, recommender models must be refreshed frequently to reflect changing preferences, seasonal effects, and catalog updates. This repeated-training requirement makes efficiency central even when vocabulary pressure is reduced through semantic IDs, retrieval pruning, or other output-space designs. In our implementation, direct item scoring remains useful for compatibility with existing evaluation and serving paths, so we focus on simple reductions to the decoding cost.

Our pre-training datasets comprise 2 trillion behavior tokens per cycle--on par with major LLM pre-training corpora, but processed on a repeated refresh cadence. Alongside standard distributed-training optimizations, we make additional changes to encoding and decoding. Below, we focus on advances in decoding efficiency.

\subsection{Efficient Decoding}

\begin{figure}
  \centering
  \includegraphics[width=0.50\textwidth]{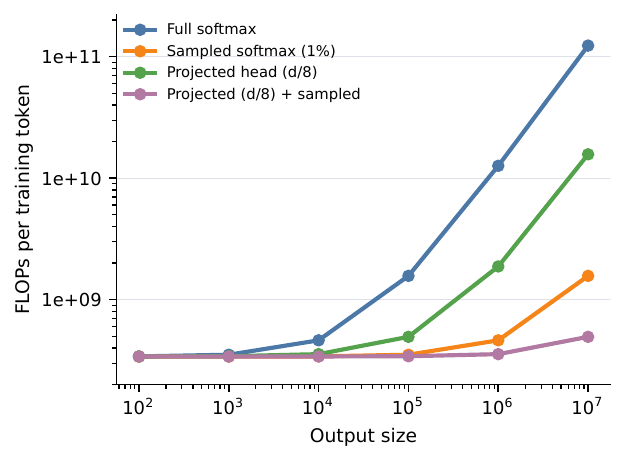}
  \caption{Estimated training FLOPs per training token for a 6-layer transformer with hidden dimension 1024 and sequence length 512. Values include a fixed backbone term plus vocabulary-dependent decoding cost; lines compare full decoding with sampled-softmax and projected-head variants as the output vocabulary grows.}
  \Description{Plot comparing estimated training FLOPs per training token across output vocabulary sizes for a 6-layer transformer with hidden dimension 1024 and sequence length 512. Values include a fixed backbone term plus vocabulary-dependent decoding cost. Lines compare full decoding, sampled softmax, projected-head decoding, and their combination.}
  \label{fig:training-flops}
\end{figure}

Even when SID or retrieval-based methods reduce the effective output space, direct item scoring remains a common and useful interface for recommendation models. For a vanilla transformer, this output layer can become a large part of training cost as the candidate set grows. Figure~\ref{fig:training-flops} illustrates this effect for a 6-layer transformer with hidden dimension 1024 and sequence length 512. The y-axis reports training FLOPs normalized by the number of training tokens, so total corpus compute scales linearly with token count. We use this analysis to motivate lightweight output-layer optimizations that are compatible with SID and retrieval-based output-space designs.

We first reduce the effective scoring set during training with sampled softmax, computing logits for positives and a uniformly sampled 1\% negative set instead of all candidate items. We reject sampled negatives that equal the target. We do not apply LogQ correction in this implementation, keeping the training recipe simple while leaving corrected sampling strategies as an orthogonal direction~\cite{khrylchenko2025logq}.

We further add a shared projected decoding head that maps the backbone hidden state from dimension $d$ to $d/8$ before item-logit computation. In production, this corresponds to a down-projection from 4096 to 512 dimensions. This targets the catalog-dependent matrix multiplication while keeping the backbone hidden dimension fixed.

At an output size of $10^6$, the illustrative vanilla configuration requires $1.26\times10^{10}$ training FLOPs per token, while sampled softmax plus a projected $d/8$ head requires $3.56\times10^8$ FLOPs per token, a $35.5\times$ reduction in this estimate. At output size $10^7$, the estimated reduction grows to roughly $249\times$. These are analytical output-layer estimates, not end-to-end training-throughput measurements.

These output-layer reductions make repeated large-model refreshes operationally feasible in our setting, rather than treating the billion-parameter model as a one-time offline artifact.

\subsection{Efficient Encoding}

Our event representation also differs from architectures that separately model heterogeneous action and feature streams. We compress the action, observation context, and item metadata for each user event into a single token-level representation. This choice keeps the sequence interface simple for downstream applications, while semantic item towers provide additional item understanding when ID embeddings are weak or unavailable (Section~\ref{sec:unseen-titles}).

\begin{figure}
  \centering
  \includegraphics[width=0.68\textwidth]{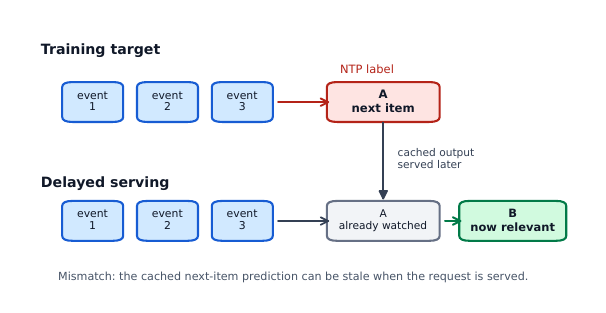}
  \caption{Latency mismatch between next-token training and delayed cached serving. Title A is the immediate next event used as the training label, but after cached serving delay it may already be consumed; title B can then become the relevant serving-time target.}
  \Description{Diagram showing next-token training over recent events where title A is the immediate target, followed by delayed serving where title A has already been watched and title B is now relevant.}
  \label{fig:latency-issue}
\end{figure}

\subsection{Efficient Development}

Finally, we use a small-to-large experiment funnel. Candidate changes are first evaluated on smaller backbones and scaled test sets designed to detect task-specific regressions. Only changes that pass this funnel are integrated into expensive billion-parameter training runs. This process is not a modeling contribution by itself, but it is important operationally: without it, the cost of validating scaling-law, decoding, MTP, and cold-start changes would be prohibitive.

\section{Multi-Token Prediction (MTP)}

\subsection{The Latency Issue in Next-Token Prediction}

Large recommenders are rarely served in a purely synchronous way. To control latency and cost, many production systems cache user embeddings, candidate lists, or pre-ranked results, refreshing them after a fixed time interval or after sufficient new activity. This pattern is related to deferred retrieval systems such as RADAR~\cite{jaspal2025radar}, but it creates a label-alignment problem for next-token training: the item immediately after the context may already be consumed by the time the cached output is used.

Figure~\ref{fig:latency-issue} illustrates the mismatch. During training, the immediate next-token label after the recent user history is title A. If the model output is cached and served later, the user may have already watched A by the time the request is served; at that point, title B can be the relevant recommendation target. A model optimized only for immediate next-token prediction can therefore place probability mass on a stale target.

\begin{figure}
  \centering
  \includegraphics[width=0.72\textwidth]{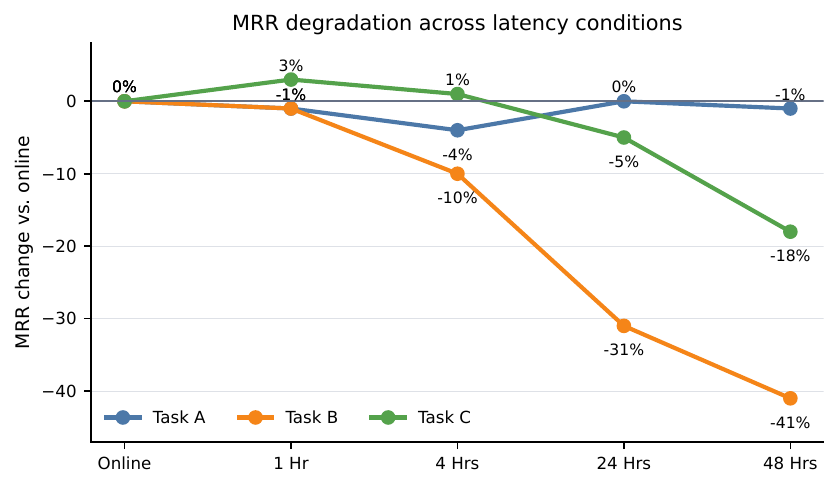}
  \caption{Relative MRR degradation as cached outputs become stale. Delays are simulated by replaying evaluation at increasing serving horizons; task definitions follow Figure~\ref{fig:scaling-law-examples} and Table~\ref{tab:task-taxonomy}.}
  \Description{Plot showing relative MRR degradation for Task A, Task B, and Task C as serving staleness increases under replayed evaluation delays.}
  \label{fig:metric-drops-staleness}
\end{figure}

Figure~\ref{fig:metric-drops-staleness} quantifies this effect by replaying evaluation under increasing serving delay. Task A is relatively robust, changing by at most 4\% in the measured range. Task B is highly latency-sensitive, dropping 31\% at 24 hours and 41\% at 48 hours. Task C is less sensitive than Task B but still drops 18\% at 48 hours. This suggests that latency alignment matters most for short-horizon tasks, while long-horizon preference tasks can tolerate stale representations better.

\subsection{Set-Valued Future Targets}

A second mismatch is that many recommendation targets are set-like. In language modeling, token order is usually semantic; in recommendation, several future items may be similarly valid and their order may be partly arbitrary. If a user is likely to watch both A and B soon, either A then B or B then A can be acceptable. Standard NTP assigns all probability mass to one observed order, penalizing other valid futures and making the objective brittle for long-term taste and exploration tasks.

We address both mismatches with multi-token prediction (MTP). For each context, we construct a weighted label set from future high-value targets and apply an exponential time decay with a one-hour half-life. The model scores the candidate set once using the same decoding head as NTP and optimizes a weighted multi-label objective over this future set. MTP therefore changes only the supervision target: at serving time, the model still scores candidates with a single decoding pass and does not require iterative generation or multiple model calls.

Formally, the loss is defined as:

\begin{equation}
  \mathcal{L} =
  - \sum_{y_i \in \mathcal{Y}}
  w_i \log p_\theta(y_i \mid x),
  \quad
  w_i = r_i \exp\left(-\ln(2)\frac{t_i-t_{\mathrm{context}}}{\beta}\right).
  \label{eq:mtp-loss}
\end{equation}

The reward weight $r_i$ can encode utility signals such as watch time, completion, freshness, or diversity. In our experiments, $\mathcal{Y}$ contains future high-value targets and $\beta$ is set to one hour, letting the model learn from future labels while emphasizing those closer to the serving state.

\begin{figure}
  \centering
  \includegraphics[width=0.78\textwidth]{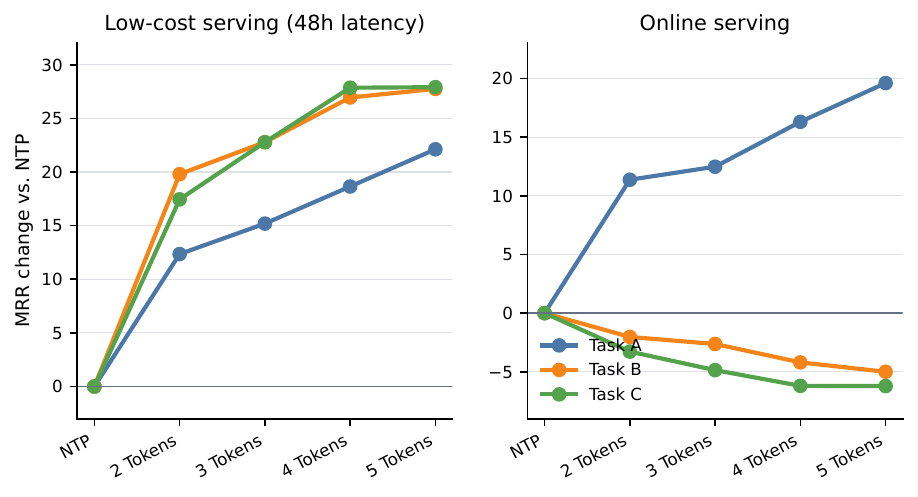}
  \caption{MTP comparison across serving scenarios. Bars report relative MRR changes for different future-target window sizes under online serving with p95 latency below one second and cached serving with a 48-hour horizon; task labels use the taxonomy in Table~\ref{tab:task-taxonomy}.}
  \Description{Plot comparing multi-token prediction performance across online serving with p95 latency below one second and cached serving with a 48-hour horizon. Bars report relative MRR changes for different future-target window sizes across Task A, Task B, and Task C.}
  \label{fig:mtp-results}
\end{figure}

Figure~\ref{fig:mtp-results} compares NTP and MTP under two serving scenarios: online serving with p95 latency below one second, and a low-cost cached path with a 48-hour serving horizon. Under the 48-hour horizon, MTP improves all three tasks, with the best five-token window yielding +22.1\%, +27.8\%, and +27.9\% relative MRR for Tasks A, B, and C respectively. Under online serving, MTP improves Task A (+19.6\% with five tokens) but reduces Tasks B and C as the window grows. These results suggest that the benefit of MTP depends on both serving delay and task type. In cached serving, the immediate next-event label may already be stale by the time the cached output is used, so training on a decayed set of future targets better matches the serving condition. In online serving, the immediate next event is still a strong target. For Tasks B and C, which depend more on fresh context, timing, and popularity signals, widening the target set can smooth away the most time-sensitive distinctions. Thus, MTP is most useful for cached or set-like targets, and should be applied more conservatively for strict online tasks.

\section{Unseen and Cold-Start Titles}
\label{sec:unseen-titles}

Cold-start titles expose a failure mode that cannot be addressed by scaling alone. Mature titles can be represented by collaborative ID embeddings learned from interaction histories, but new or sparse titles provide little evidence for that channel. Prior work addresses related cold-start problems through semantic IDs, content-derived tokenization, or exploration policies~\cite{rajput2023recommender,zhu2024cost,mei2025semantic,wang2025item}. Our design is complementary: we keep collaborative ID embeddings for mature titles, while adding multimodal semantic item towers so the same generative decoder can score titles from content and metadata when ID evidence is weak.

\begin{figure}[!t]
  \centering
  \includegraphics[width=0.58\textwidth,trim={0.10in 0.20in 0.10in 0.17in},clip]{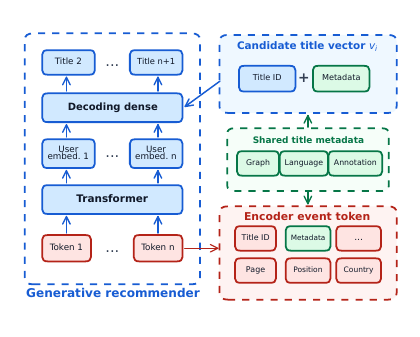}
  \caption{Shared semantic title metadata for encoder events and decoder title representations. Blue denotes learned representations and computation, red denotes input-side event context, and green denotes reusable title metadata from knowledge-graph message passing, LLM2Vec language embeddings, and human annotation features.}
  \Description{Single-column conceptual architecture diagram. The left side shows event tokens passed through a transformer, user embeddings, dense decoding, and title scoring. The right side places the candidate title vector above the encoder event token and uses three shared metadata boxes labeled Graph, Language, and Annotation. Green metadata boxes appear in both the candidate-vector construction and the encoder event token; contextual fields and an ellipsis for additional fields appear only in the encoder event token.}
  \label{fig:semantic-tower}
  \vspace{-0.5em}
\end{figure}

Figure~\ref{fig:semantic-tower} illustrates the resulting separation between encoder-side event representation and decoder-side title scoring. The left side shows the standard generative path: event tokens are encoded by the Transformer, converted into user representations, and passed to a dense decoding layer. On the right, semantic title metadata is shared by two consumers. These metadata sources include graph features learned by message passing over the title knowledge graph, language embeddings produced with LLM2Vec~\cite{behnamghader2024llm2vec}, and editorial annotation features derived from human labels. They enrich the encoder event token when the title appears in a user history, and they also contribute to the candidate title vector scored by the decoder. Contextual fields such as page, position, and country remain input-side event features; they condition the user representation but are not part of candidate-title vectors. The color coding mirrors this distinction: blue denotes learned model representations and computation, red denotes encoder-side event/context fields, and green denotes reusable title metadata.

At decoding time, semantic information enters through the title representation used by the dense output head. Let $h_u$ denote the user representation produced by the Transformer, let $\mathcal{V}$ denote the item vocabulary used by the model, and let $v_i$ denote the candidate title vector for title $i$. The decoder score is computed as:
\begin{equation}
\begin{aligned}
  z_i &= \phi_{\mathrm{sem}}(e_i^{\mathrm{graph}}, e_i^{\mathrm{lang}}, e_i^{\mathrm{ann}}), \\
  \tilde{e}_i^{\mathrm{ID}} &=
  \begin{cases}
    e_i^{\mathrm{ID}}, & i \in \mathcal{V}, \\
    e^{\mathrm{OOV}}, & i \notin \mathcal{V},
  \end{cases} \\
  v_i &= \psi_{\theta}(\tilde{e}_i^{\mathrm{ID}}, z_i), \\
  s(u, i) &= g_{\theta}(h_u)^\top v_i .
\end{aligned}
\end{equation}
Here, $e_i^{\mathrm{graph}}$, $e_i^{\mathrm{lang}}$, and $e_i^{\mathrm{ann}}$ are the Knowledge Graph, Language Embedding, and Annotation feature embeddings for title $i$, and $\phi_{\mathrm{sem}}$ maps them into the semantic title representation $z_i$. The term $e_i^{\mathrm{ID}}$ is the collaborative ID embedding for an in-vocabulary title, $e^{\mathrm{OOV}}$ is the learned out-of-vocabulary embedding, and $\tilde{e}_i^{\mathrm{ID}}$ is the ID-side embedding actually used by the decoder. The learned functions $\psi_{\theta}$ and $g_{\theta}$, with parameters $\theta$, construct $v_i$ and project $h_u$ into the decoding space; $(\cdot)^\top$ denotes the inner product used for scoring; and $s(u,i)$ is the resulting user-title score. In-vocabulary titles use both $e_i^{\mathrm{ID}}$ and $z_i$. Cold-start titles that do not exist in $\mathcal{V}$ are assigned the learned OOV embedding and rely on $z_i$ for title-specific evidence. This separates stable title evidence from transient serving context: semantic metadata can define candidate-title vectors, while page, position, country, and other contextual fields only condition the user-side representation.

The key training idea is to expose the decoder to both mature-ID and weak-ID regimes, so it learns to combine collaborative ID evidence with semantic title evidence rather than relying on either source in isolation. During pre-training, we randomly replace either the input-side or output-side collaborative item embedding with the learned OOV embedding, using a masking probability aligned with the measured online cold-start rate. These masked examples match the serving path for titles outside $\mathcal{V}$. They train the decoder to score a title from the learned OOV embedding plus semantic evidence, so the semantic tower carries title-specific information when the item ID is unavailable.

\begin{table}
  \centering
  \caption{Main differences between the baseline and scaled production-shadow models.}
  \label{tab:combined-system}
  \begin{tabular}{@{}p{0.23\linewidth}p{0.25\linewidth}p{0.38\linewidth}@{}}
    \toprule
    Dimension & Baseline & Scaled model \\
    \midrule
    Backbone & 2M parameters & 1B parameters \\
    Decoding & Full softmax & Sampled softmax + projected head \\
    Objective & Next-token prediction & Multi-token prediction \\
    Cold start & ID embeddings & Semantic towers + ID masking \\
    \bottomrule
  \end{tabular}
  \Description{The table compares the main modeling differences between the baseline and scaled model used in production-shadow evaluation.}
\end{table}

\section{Production-Shadow Evaluation}
\label{sec:results}

\subsection{Setting}

The preceding sections evaluate individual design questions through scaling-law fits, analytical cost estimates, serving-delay replay, and MTP comparisons. The production-shadow comparison evaluates a scaled production system rather than a scale-only ablation. Table~\ref{tab:combined-system} summarizes the main differences between the 2M-backbone baseline and the 1B-backbone model.

\begin{figure}[!ht]
  \centering
  \includegraphics[width=0.68\textwidth]{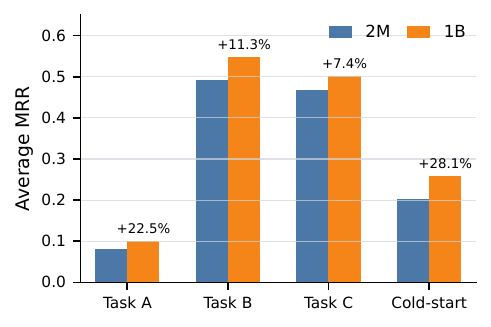}
  \caption{Weekly production-shadow MRR over 1M users. The 1B-backbone model is compared with a 2M-backbone baseline, excluding embedding and decoding layers from the size count, across the three task categories and cold-start titles outside the model vocabulary.}
  \Description{Bar chart reporting weekly average MRR for a 1B-backbone model and a 2M-backbone baseline across Task A, Task B, Task C, and cold-start titles, computed from production-shadow traffic over 1M users.}
  \label{fig:weekly-performance}
\end{figure}

We run the candidate model in production-shadow traffic for 1M users over a one-week window. For each shadowed request, downstream systems consume the model through the same integration path used for production recommendation, but the candidate model does not affect the user experience. We log the shadow outputs and evaluate them against downstream labels collected during the same window. This setting evaluates whether the foundation model produces useful representations and scores across downstream integrations, rather than measuring the causal lift of a single product surface.

\subsection{Results}

Figure~\ref{fig:weekly-performance} compares a 1B-backbone model with a 2M-backbone baseline, excluding embedding and decoding layers from the size count, in this 1M-user shadow-traffic evaluation. The 1B-backbone model improves MRR across all reported slices: +22.5\% for Task A, +11.3\% for Task B, +7.4\% for Task C, and +28.1\% for cold-start titles. The largest gain on cold-start titles supports the semantic-tower design in Section~\ref{sec:unseen-titles}. The task-level variation in gains is consistent with the scaling-law analysis, which suggests that scaling decisions should account for both fitted headroom and task importance. We interpret these results as broad production-shadow evidence that the 1B-backbone model transfers across downstream integrations. Consistent with this shadow evaluation, downstream integrations of the model have also produced positive outcomes in multiple production A/B tests.

\FloatBarrier

\section*{Acknowledgments}
We thank Baolin Li, Bella Nicholson, Dan Zheng, Dhaval Patel, Divya Gadde, Michael Tu, Sejoon Oh, Sudarshan Lamkhede, Wei Wang, Yesu Feng, and Zhe Zhang for their contributions to the modeling, infrastructure, evaluation, and production integration work that made this system possible. We also thank the anonymous reviewers for their valuable comments and suggestions, which helped improve the clarity and presentation of this paper.

\bibliographystyle{plain}
\bibliography{references}

\clearpage
\appendix

\end{document}